\documentclass[preprint,amsmath,amssymb,floatfix]{revtex4-1}
\usepackage{graphicx}
\usepackage{dcolumn}
\usepackage{bm}
\usepackage{upgreek}
\usepackage[mathlines]{lineno}

\begin{document}
\title{Dual-wavelength vortex beam with high stability in diode-pumped Yb:CaGdAlO$_4$ Laser}

\author{Yijie Shen}
\author{Yuan Meng}
\author{Xing Fu} 
\author{Mali Gong} 
\email{shenyj15@mails.tsinghua.edu.cn}
\affiliation{State Key Laboratory of Precision Measurement Technology and Instruments, Department of Precision Instrument, Tsinghua University, Beijing 100084, China}

\date{\today}

\begin{abstract}
We present a stable dual-wavelength vortex beam carrying orbital angular momentum (OAM) with two spectral peaks separated by a few terahertz in diode-pumped Yb:CaGdAlO$_4$ (CALGO) laser. The dual-wavelength spectrum is controlled by the pump power and off-axis loss in laser resonator, arising from the broad emission bandwidth of Yb:CALGO. The OAM beam is obtained by a pair of cylindrical lens that serves as an $\pi/2$ convertor for the high-order Hermite-Gaussian modes. The stability is verifed that the $1\hbar$ OAM beam with two spectral peaks at 1046.1~nm and 1057.2~nm (3.01~THz interval) can steadily operate for more than three hours. It has great potential for scaling the application for OAM beams in Terahertz spectroscopy, high-resolution interferometry, and so on.


\end{abstract}

\maketitle

\section{Introduction}
In dual-wavelength spectroscopic technology, the dual-wavelength lasers with frequencies at two peaks separated by a few terahertz (THz) are widely used in various applications such as THz imaging and spectroscopy \cite{1,2}, high-resolution interferometry \cite{3}, optical and sensing \cite{4,5}, etcetera. 

Recently, the optical vortex beams carrying orbital angular momentum (OAM) attract intensive research interest, for its broad applications in optical tweezers \cite{6}, optical communications \cite{7,8,9}, quantum entanglement \cite{10,11} utilizing its unique properties of helical phase. Furthermore, the dual-wavelength OAM beams are beneficial for the advanced applications, due to its combinative characteristics of both helical phase and dual-wavelength spectrum. A dual-wavelength OAM beam based on Nd:Lu$_2$O$_3$ laser was previously reported \cite{a1}, but the two spectral peaks are unbalanced and unadjustable. Hence, the optimization of the spectral profile and improving the tunability for matching the modern spectroscopy applications are still required. As a new crystal with ultra-broad emission band, Yb:CaGdAlO$_4$ (CALGO) shows outstanding performance to satisfy these requirements \cite{12,13,14,15,16}. Our group has recently demonstrated a wavelength-tunable vortex beam with the highest order of $15\hbar$ OAM\cite{a2}, however, the conditions for high-stability dual-wavelength emitting have never been investigated.

In this work, we demonstrate a dual-wavelength OAM beams with two stable spectral peaks separated by a few THz. By combining the feature of broad emission band of Yb:CALGO and special coating of cavity mirrors, the dual-wavelength generation in a diode-pumped solid-state laser (DPSSL) can be effectively obtained. Using a $\pi/2$ convertor \cite{17}, the vortex beam is converted from the Hermite-Gaussian (HG) mode produced by off-axis pumping. Under some certain pump powers and off-axis displacements, the stable OAM beams with dual-wavelength spectrum can be generated. We experimentally verified the spectral stability that a $1\hbar$ OAM beam with two spectral peaks of 1046.1~nm and 1057.2~nm (separated by 3.01~THz) can operate for more than three hours.
 
\section{Experimental Design}
\begin{figure*}
	\centering
	\includegraphics[width=\linewidth]{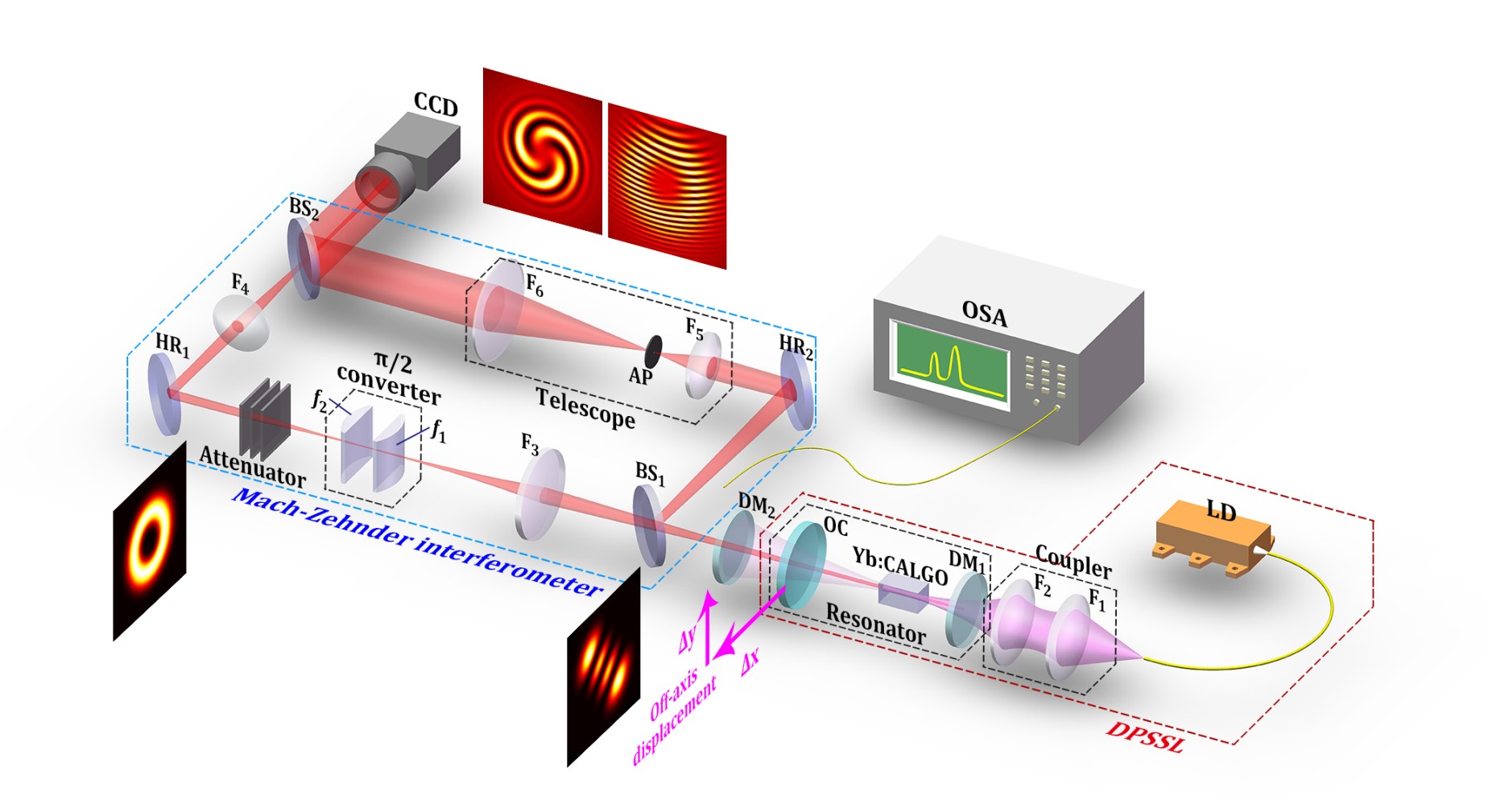}
	\caption{\label{f1} Schematic of the experimental setup: LD, laser diode; OSA, optical spectral analyzer; CCD, charge coupled device; DM, dichroic mirror; OC, output coupler; HR, high-reflective mirror; AP, aperture; DPSSL, diode-pumped solid-state laser.}
\end{figure*}

The experimental setup, as depicted in Fig.~\ref{f1}, includes two main parts: the off-axis-pumped DPSSL for generating high-order HG mode and the Mach-Zehnder interferometer for generating and measuring the OAM.

\begin{figure}
	\centering
	\includegraphics[width=0.6\linewidth]{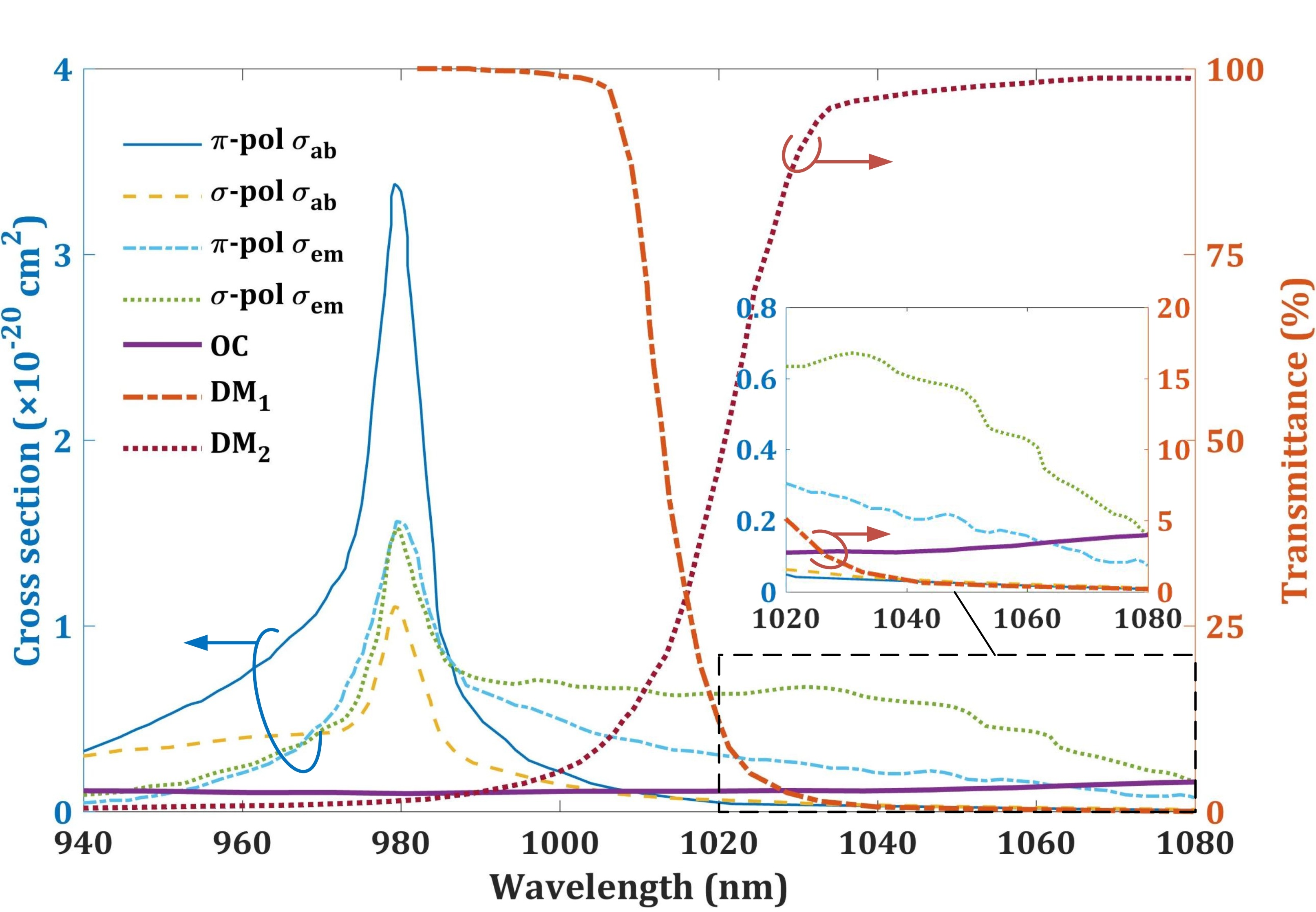}
	\caption{\label{f2} The polarization-dependent absorption and emission cross sections of Yb:CALGO and the transmittance curves of our DM$_1$, DM$_2$, and OC mirrors.}
\end{figure}

For the DPSSL, a $2\times2\times4$~mm$^3$ (4 mm is along laser direction) a-cut 5~at.\%-doped Yb:CALGO with two end surfaces antireflective (AR) coated for laser and pump light was used as gain medium, which was wrapped with the indium foil and conductively water cooled at the temperature of 18~$^\circ$C. The laser is generated by a linear plano-concave resonator with a concave output coupling mirror (OC, radius of curvature of 300~mm) and a flat dichroic mirror [DM$_1$, high-reflection (HR) coated for laser and AR coated for the pump light]. Based on the polarization-dependent emission spectra (Fig.~\ref{f2}: the $\sigma$-polarization is superior at 1000-1080~nm), the laser was  $\sigma$-polarized due to the gain competition. The crystal was pumped by a 976~nm fiber-coupled laser diode (LD) (Han’s TCS, core: 105~$\upmu$m, NA: 0.22, highest power: 110~W) with the pump waist radius of 200 $\upmu$m through a coupler including two identical AR-coated convex lenses ($F_1=F_2=60$~mm) and DM$_1$. The DM$_2$ (AR coated for laser and HR coated for the pump light) was used to filter residual pump light. 

For the interferometer, the two arms were formed by two beam splitters (BS$_1$ and BS$_2$, 45$^\circ$ incidence, T:R=1:9) and two 45$^\circ$ HR mirrors (HR$_1$ and HR$_2$). For the first arm, the laser was incident into the $\pi/2$ convertor after being focused by a convex lens ($F_3=180$~mm) and converted into Laguerre-Gaussian (LG) mode:
\begin{align} 
LG_{p,l}(r,\phi,z)=& \sqrt{\frac{2p!}{\pi\left(p+\left|l\right|\right)!}}\frac{1}{w(z)}\left[\frac{r\sqrt{2}}{w(z)}\right]^{\left|l\right|}\exp\left[\frac{-r^2}{w^2(z)}\right]\notag \\ &L_p^{\left|l\right|}\left(\frac{2r^2}{w^2(z)}\right)\exp\left({\rm i}l\phi\right)\exp\left[\frac{{\rm i}\pi r^2z}{\lambda\left(z^2+z_R^2\right)}\right]\notag \\ &\exp\left[-{\rm i}\left(2p+\left|l\right|+1\right)\tan^{-1}\left(\frac{z}{z_R}\right)\right],
\end{align}
where $r=\sqrt{x^2+y^2}$, $\phi=\tan^{-1}(y/x)$, the $1/{\rm e}$ radius of the Gaussian term is given by $w(z) = w(0)\sqrt{(z^2+z_R^2)/z_R^2}$ with $w(0)$ being the beam waist, $z_R$ is the Rayleigh range, and $L_p^{|l|}(\cdot)$ is an associated Laguerre polynomial; the produced LG$_{0,l}$ beam with OAM of $l\hbar$ passed through the attenuator and a convex lens ($F_4=150$~mm) and was captured by a CCD camera (Spiricon, M2-200s) after being focused by a convex lens ($F_4=150$~mm); the $\pi/2$ convertor includes two identical convex-plane cylindrical lenses ($f=f_1=f_2=25$~mm) with separation of 35.4~mm ($\sqrt{2}f$); the transmittance of the attenuator can be adjusted by changing the filters. For the second arm, a confocal telescope including two convex lenses ($F_5=60$~mm, $F_6=300$~mm) with an aperture was used to convert the laser into a near plane wave, which was captured by the CCD and formed the interference pattern:
\begin{equation}
I=\left|LG_{0,l}(r,\phi,z)+\eta LG_{0,0}(r,\phi,\infty)\exp\left[{\rm i}\frac{\theta_xx+\theta_yy}{\lambda}\right]\right|^2,
\label{e2}
\end{equation}
where $\eta$ is the intensity ratio of two beam and $\theta_x$ ($\theta_y$) represents the inclined angle at horizontal (vertical) direction.

The laser spectrum was measured by an optical spectrum analyzer (Agilent, 86140B). The laser power was measured by a thermopile power-meter (Ophir, FL250A-LP1-DIF).

Thanks to the broad and flat emission band (approximate 80~nm) of Yb:CALGO \cite{13,14}, it is possible to directly generate broad-band wavelength-tunable and dual-wavelength lasers in oscillators \cite{15,16}. A broad-band coating on cavity mirrors is elaborately designed to provide enough longitudinal mode gain competition for realizing the wavelength-tunable property. As the HR mirror of the laser resonator, DM$_1$ was AR coated at 976~nm and HR coated at 1040-1080~nm. The output transmittance of OC was about 2\% at 1030-1080~nm, among which a slight rise existed as the wavelength increases. The absorption and emission cross sections of Yb:CALGO and the transmittance curves of DM$_1$, DM$_2$, and OC mirrors are shown in Fig.~\ref{f2}.

\section{Results and discussions}
\subsection{Dual-wavelength emission property of Yb:CALGO}
\begin{figure}
	\centering
	\includegraphics[width=0.6\linewidth]{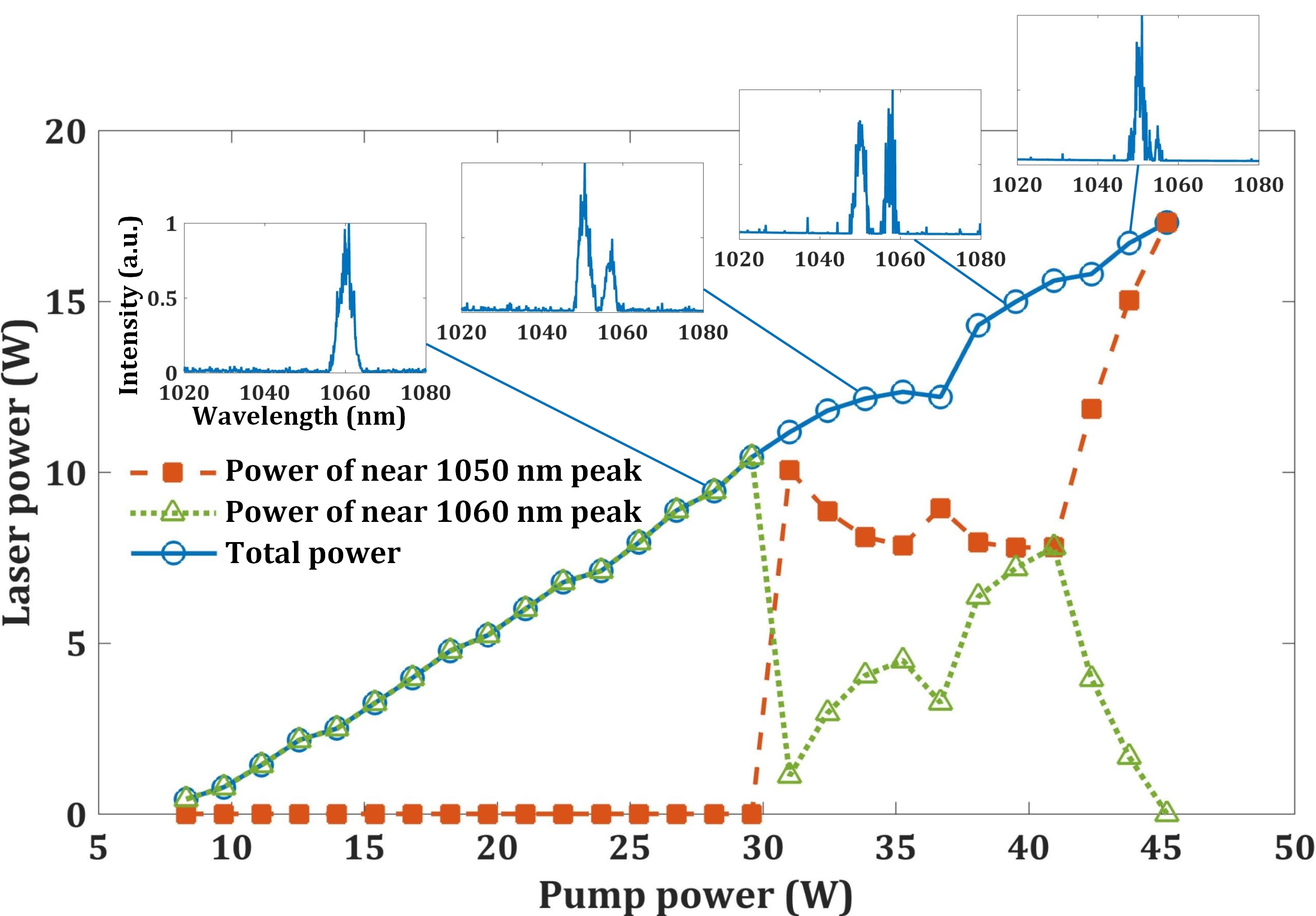}
	\caption{\label{f3} The laser power and spectrum evolution versus the pump power when the cavity and pump light were strictly coaxial ($\Delta r=0$). Inserts: corresponding measured spectrum profiles in the evolution.}
\end{figure}

Considering the broad emission spectrum with plateau profile of Yb:CALGO and the broadband coating design used in our experiment, the strong gain competition makes it possible that there are two superior longitudinal modes for forming a dual-wavelength spectrum. As expected, we successfully observed the dual-wavelength emission under some pump powers. When the cavity and pump light were strictly coaxial, the power and spectral evolution is depicted in Fig.~\ref{f3}. The pump threshold of the laser oscillator was around 8.3~W. The output spectrum maintained the single-peak profile until the pump power increased to 32.1~W, at which a dual-wavelength was observed. Afterwards, we continuously increased the pump power from 32.1~W to 46.8~W, and the dual-wavelength oscillation was maintained while the spectral intensity of two peaks varied versus the pump power. 

\subsection{Generation of dual-wavelength vortex beam}
\begin{figure}
	\centering
	\includegraphics[width=1.1\linewidth]{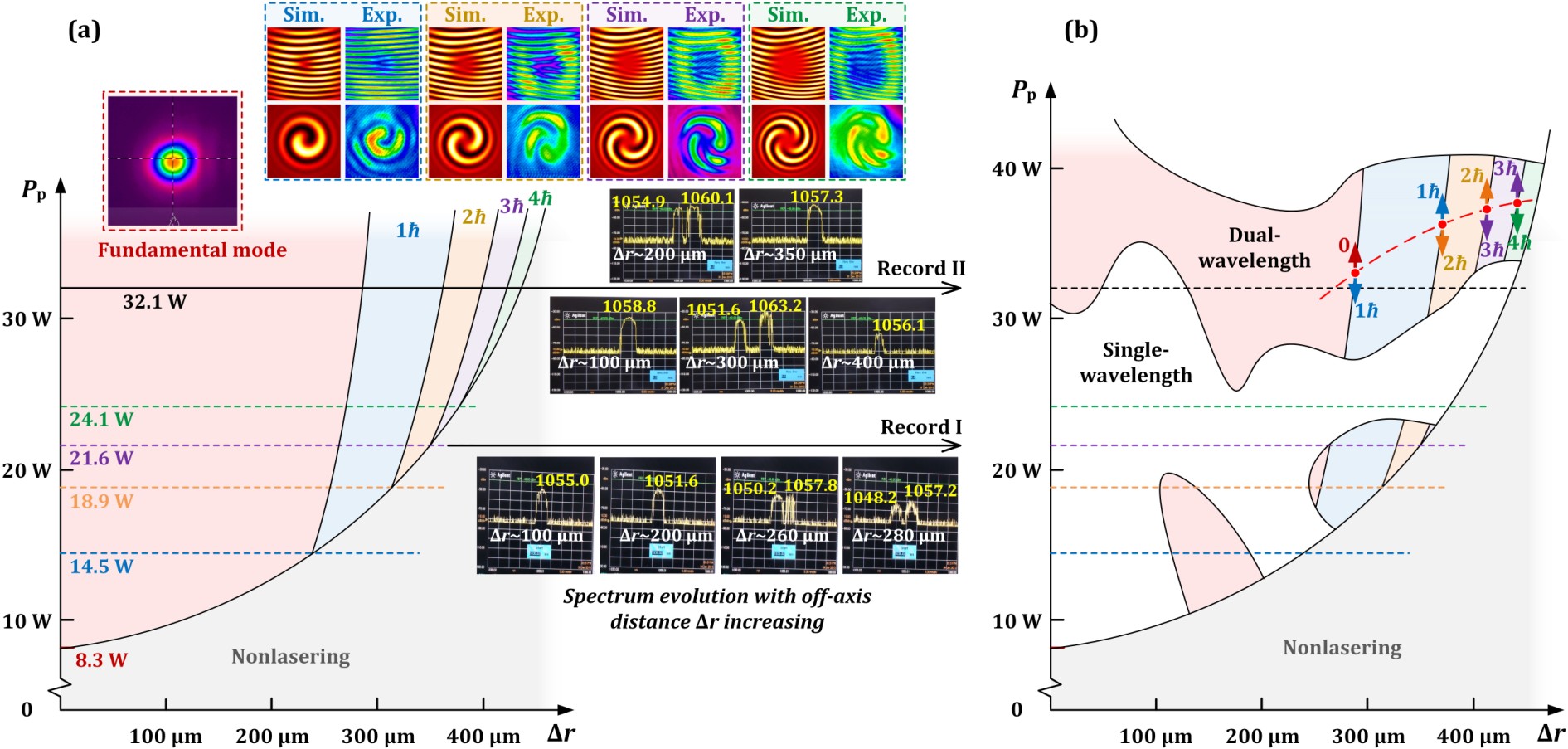}
	\caption{\label{f4} $P_{\rm p}$-$\Delta r$-mode map: (a) the principle of mode evolution with pump power and off-axis distance. The upward insets: the theoretical and experimental interference patterns to verify the vortex beam with various OAMs. The right insets: The spectrum evolution with off-axis distance increasing under the fixed pump powers at 32.1 W and 21.6 W, noted as \textbf{Record I} and \textbf{Record II}; (b) the single-wavelength and dual-wavelength regions; the thick dash line marks a track that the OAM can be tuned maintaining the dual-wavelength spectrum; at the dot positions, the OAM can be tuned to the adjacent value by adjusting the pump power.}
\end{figure}

In our experiment, the vertical and horizontal off-axis distance of OC, $\Delta x$ and $\Delta y$, can be precisely adjusted to generate the high-order HG mode along an inclined direction. A HG$_{0,l}$ mode placed along the 45$^\circ$ diagonally direction can be directly converted into $l\hbar$-OAM beam (LG$_{0,l}$ mode) via a pair of cylindrical lens \cite{17}. The output mode of the DPSSL is mainly dependent on the pump power ($P_{\rm p}$) and off-axis distance $\Delta r=\sqrt{\Delta x^2+\Delta y^2}$. According to our experimental results, we depicted the $P_{\rm p}$-$\Delta r$-mode map to reveal the principle of mode evolution shown in the Fig.~\ref{f4}(a). The pump threshold is 8.3~W with TEM$_{00}$ output when the cavity is strictly coaxal. If we increase the pump power to 14.5~W and adjust the OC with the off-axis distance about 250~$\upmu$m, the $1\hbar$-OAM beam can be generated. Similarly, continuously increased the pump power and off-axis distance to (18.9~W, 320~$\upmu$m), (21.6~W, 350~$\upmu$m), and (24.1~W, 380~$\upmu$m) can lead to the generation of $2\hbar$, $3\hbar$, and $4\hbar$ OAMs respectively. The interference patterns to verify the vortex beam with various OAMs are theoretically and experimentally demonstrated in Fig.~\ref{f4}(a). The vortex patterns are obtained when the two interference beams are coaxial, i.e. $\theta_x=\theta_y=0$ in Eq.~(\ref{e2}). When we fixed the pump power at 21.6~W and continuously increased the off-axis distance until the nonlasering state, the dual-wavelength spectrum can be obtained accordingly with the generation of OAM, as shown in \textbf{Record I} in Fig.~\ref{f4}(a). We noted that the $1\hbar$- and $2\hbar$-OAM beams in this process can overlap with dual-wavelength region. When we fixed the pump power at 32.1~W, similar evolution can also be observed, the spectrum first changed from single-peak shape to dual-wavelength shape and then can return to the case of single-peak but with a different center wavelength, as shown in \textbf{Record II} in Fig.~\ref{f4}(a). After many times recording the spectral evolutions versus off-axis distant at different pump powers, we obtained the overlapped region between the OAM beam states and dual-wavelength spectrum states, as shown in Fig.~\ref{f4}(b), which reveals that the corresponding dual-wavelength OAM beam can be produced. As can be seen, there is a dual-wavelength region that can cover from $1\hbar$ to $4\hbar$ OAM states, in which the OAM can be tuned while the spectrum can simultaneously maintain the dual-wavelength shape by properly controlling pump power and off-axis distance. 

\begin{figure}[htbp]
	\centering
	\includegraphics[width=0.7\linewidth]{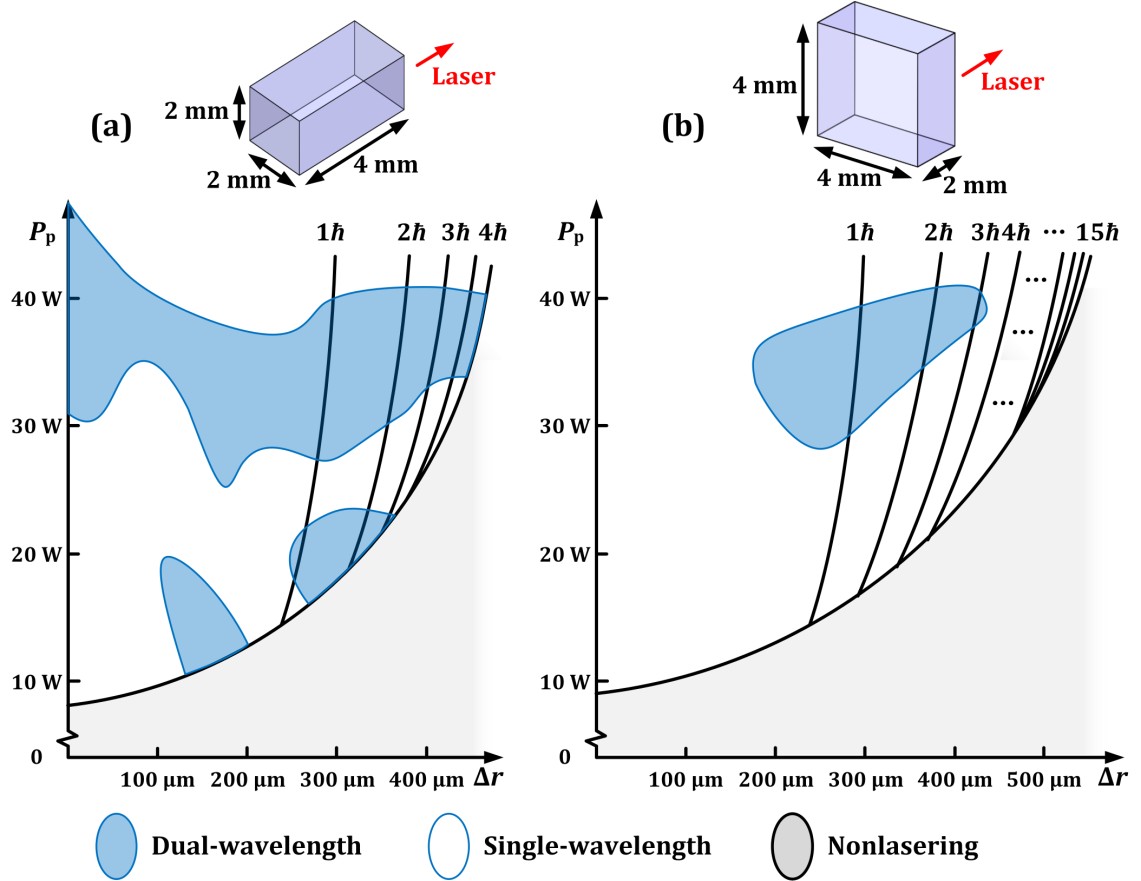}
	\caption{\label{f5} The comparison of single- and dual-wavelength regions in $P_{\rm p}$-$\Delta r$-mode map using crystals of different sizes: (a) 2$\times$2$\times$4 mm$^3$ (4 mm is along the laser direction) and (b) 4$\times$4$\times$2 mm$^3$ (2 mm is along the laser direction).}
\end{figure}

For the mechanism of the high-order modes generation in off-axis cavity, the physical origin can be related to the effect of fractional degeneracy and the emergence of ray-wave duality \cite{18}, which reveals the reason of OAM dependence on off-axis distance in our system. The laser spectrum is induced by the longitudinal mode gain competition. Specially, the Yb:CALGO has a broad and flat emission spectrum \cite{13,14,15}, which means the gain competition is very sensitive to the intracavity loss and the strong gain competition may easily lead to dual-wavelength spectrum \cite{16}. Therefore, the crystal size is an important parameter to influent the dual-wavelength vortex beam emitting, which is related to the laser gain and aperture effect. In our previous work \cite{a2}, we use a 4$\times$4$\times$2 mm$^2$ crystal (flat-sheet shape) to scaling the maximum OAM to $15\hbar$, because the larger aperture provide more potential of ray-wave duality rather than 2$\times$2$\times$4 mm$^2$ crystal (thin-rod shape). However, the flat-sheet shaped crystal is to the disadvantage of stable dual-wavelength vortex beam emitting because the gain length is too short to provide strong gain competing. Fig.~\ref{f6}(a,b) shows the results of single- and dual-wavelength regions in $P_{\rm p}$-$\Delta r$-mode maps for the case of using the 2$\times$2$\times$4 mm$^2$ and 4$\times$4$\times$2 mm$^2$ crystals respectively. Through the comparative results, the dual-wavelength region for the case of using 4$\times$4$\times$2 mm$^2$ crystal is much smaller than that of 2$\times$2$\times$4 mm$^2$ case. We only observed a limited region that can overlap dual-wavelength state and OAM state. Therefore, the condition for generating stable dual-wavelength emitting is different from that of generating stable high-order modes. Using flat-sheet shaped crystal can effectively enlarging the OAM-tunable region, while using the thin-rod shaped crystal can be beneficial to obtain stable dual-wavelength vortex beam.

\subsection{Stability of dual-wavelength vortex beam}
\begin{figure}[htbp]
	\centering
	\includegraphics[width=0.6\linewidth]{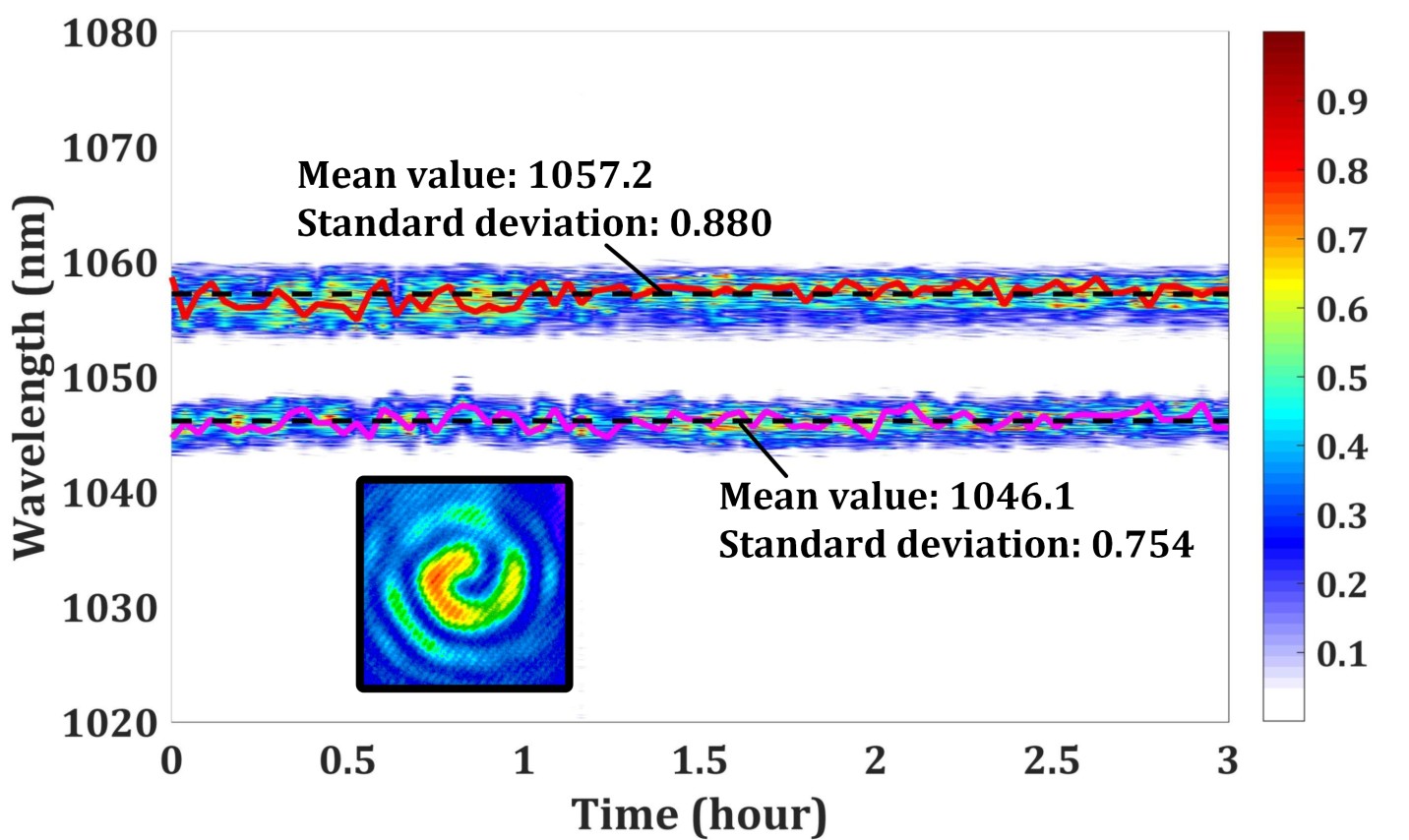}
	\caption{\label{f6} The stability verification: the recorded dual-wavelength spectrum evolution for three hours of the $1\hbar$-OAM beam under the pump power of 21.6~W and the off-axis distance about 300~$\upmu$m.}
\end{figure}

To test the stability of the dual-wavelength vortex beams, the spectrum of the $1\hbar$-OAM vortex beam ($P_{\rm p}\sim$35.0 W, $\Delta r\sim$300 μm) was recorded for more than 3 hours, as shown in Fig.~\ref{f6}. The dual-wavelength spectrum is well maintained with one peak at 1057.2~nm and the other one at 1046.1~nm and the intensities of two peaks were approximately equal. The standard deviations of the two center wavelengths were 0.881~nm and 0.754~nm respectively. The profile of two spectral peaks were very balanced and the two center wavelength were separated by 11.1~nm (corresponding to 3.01~THz), which are satisfactory for applications of dual-wavelength spectroscopic techniques.

\section{Conclusion}
In conclusion, we demonstrate a stable dual-wavelength vortex beams generated from the Yb:CALGO DPSSL with a $\pi/2$ convertor. The dual-wavelength spectrum and OAM can be flexibly controlled by the off-axis distance and pump power due to the broad-emitting-band property of Yb:CALGO and coating design. We depicted the $P_{\rm p}$-$\Delta r$ map where the overlapped region between dual-wavelength spectrum and OAM beams reveals the generation of corresponding dual-wavelength OAM beams. We experimentally tested the stability of the dual-wavelength vortex beam that the 1$\hbar$-OAM beam with two spectral peaks at 1046.1~nm and 1057.2~nm (separated by 3.01~THz) which are capable of steadily operating for more than three hours. The dual-wavelength OAM beams separated by a few THz possesses great potential for scaling the applications for OAM beams such as Terahertz spectroscopy, high-resolution interferometry.

\end{document}